Accepted Version

Publication date: February 2023
Embargo: 24 Months
European Union, Horizon 2020, Grant Agreement number: 857470 — NOMATEN — H2020-WIDESPREAD-2018-2020
DOI: https://doi.org/10.1016/j.apsusc.2022.155465

# Mechanical behavior of ion-irradiated ODS RAF steels strengthened with different types of refractory oxides

M. Frelek-Kozak[1*], Ł. Kurpaska[1], K. Mulewska[1], M. Zieliński[1], R. Diduszko[1,2], A. Kosińska[1], D. Kalita[1], W. Chromiński[1,3], M. Turek[4], K. Kaszyca[2], A. Zaborowska[1], J. Jagielski[1,2]

1. NOMATEN Centre for Excellence - National Centre for Nuclear Research, A. Soltana 7 St., 05-400 Otwock, Poland
2. Łukasiewicz Research Network – Institute of Microelecotronic and Fotonics, Wólczyńska 133 St., 01-919 Warsaw, Poland
3. Warsaw University of Technology, Faculty of Materials Science and Engineering, st. Wołoska 141, 02-507 Warsaw, Poland
4. Maria Curie-Skłodowska University, M. Curie-Skłodowskiej Sq. 5, 20-031 Lublin, Poland

* corresponding author: malgorzata.frelek-kozak@ncbj.gov.pl

**Abstract**

One of the most promising candidates for constructing IV generation nuclear reactors is Oxide Dispersed Strengthening (ODS) Reduced Activation Ferritic (RAF) steel. It is known that introducing refractory oxides to the ferritic matrix makes it possible to obtain a brand-new kind of materials with an excellent set of properties. In the present work, the authors focused on verifying materials' structural and mechanical properties strengthened by three different types of refractory oxides submitted to ion-irradiation. Three materials with an elemental chemical composition of 12%Cr, 2%W, 0.3%Ti, and strengthened with 0.3% $Y_2O_3$ or $Al_2O_3$ or $ZrO_2$ were produced by mechanical alloying and subsequently consolidated by Spark Plasma Sintering technique. Manufactured materials have been submitted to high energy (500 keV) Ar-ion irradiation at room temperature with three fluences (up to $5x10^{15}$ ions/cm$^2$). This procedure allowed to generate a thin, strongly damaged zone with approximate thickness of 230 nm. SEM/EBSD and TEM bservations, Grazing Incidence X-ray Diffraction analysis, and nanoindentation tests have been included in examination of modified layers. Implementation of these techniques allowed to reveal alteration of structural and mechanical features between unmodified material and radiation-affected layer. Obtained results showed a strong correlation between the strengthening oxide and materials' microstructural and mechanical behavior under radiation damage. It has been proved that below $1x10^{15}$ ions/cm$^2$ mechanical properties in the modified layer of all materials are very similar. Reported behavior may be related to the efficient annealing of the radiation defect process. Above this limit, significant differences between the materials are visible. It is believed that described phenomenon is directly related to the presence of the structural features and their capacity to act as defect sinks. Consequently, type of dominant mechanisms occurring in modified layer is proposed.

1. **<u>Introduction</u>**

Currently, ongoing works on IV generation of nuclear reactors focus on several aspects of its exploitation: (i) excel in safety, (ii) proliferation resistance, (iii) sustainability and economics, and finally, (iv) novel opportunities to use nuclear power in industry (such as cogeneration of process heat or production of synthetic fuel or hydrogen) [1,2]. These design assumptions make requirements for constructional materials rigorous and demanding. The higher temperature of work (up to 600°C), corrosive coolants, and exploitation time extended up to 80 years are new challenges that must overcome materials dedicated to IV generation of nuclear installations [2,3]. Due to its excellent mechanical properties at wide range of temperatures [4], high radiation resistance [5], and ability to utilization in corrosive coolant [6], oxide dispersion strengthened (ODS) steels have a high potential to be used as a structural component of fast breeder reactors [7]. This unique set of properties is related to fine stable refractory oxides homogenously dispersed in the material's structure. The most common introduced oxide is yttria ($Y_2O_3$), which - during production process - interacts with other elements and creates more complex forms, such as pyrochlore $Y_2Ti_2O_7$ [8] or hexagonal $Y_2TiO_5$ [9]. Its beneficial influence on the functional properties of steel has been widely proved in the literature [2,10–12].

Surprisingly, the impact of other types of refractory oxides is still poorly investigated. Only partial information published by two independent groups from the Czech Academy of Science and German KIT [13–15] can be found in the literature. For example, Siska *et al.* [15] suggested that alumina particles provide the highest strengthening effect. This has been explained by the stress decrease around the particles, which is induced by a high difference between Young's moduli of alumina and the ferritic matrix. This stress field can trap the dislocation in the vicinity of particles. Stress field related to mismatch of elastic constants inhibits dislocation movement at temperature range up to 600°C. A different approach has been presented by Hoffman *et al.* [14], who studied four different oxides (MgO, $La_2O_3$, $Ce_2O_3$, and $ZrO_2$). These were selected by looking at their thermal stabilities and Gibbs-free enthalpies of various chemical compositions. It has been proved that some properties of the MgO-added alloys display better mechanical properties than the reference $Y_2O_3$-containing material. Mechanical tests show that the selected oxides are potential candidates to produce oxide dispersion-strengthening ferritic steels. Some alloys strengthened with alternative oxides (MgO, $Ce_2O_3$) performed better than the reference material in Charpy impact tests. $Y_2O_3$-added alloys displayed the best performance among all the investigated materials regarding tensile behavior. This is due to the excellent high-temperature stability and more uniform size distribution of the oxide particles. In conclusion, during the last decades, information about steel strengthened by alternative oxides has not been fully understood, which is the prime motivation of this study. Information about properties of such steel under radiation damage are even more scarce [16]. It is known that ODS steels may suffer very high radiation damage (exceeding even 150 dpa). Since no information regarding radiation damage build-up behavior exists, we propose this study focus on the early stage of damage accumulation. This stage significantly affects migration and defect recovery at high dpa. This point constitutes the second motivation of this work. Introduction of yttria into ferric matrix is directed by high thermal stability and low diffusion rate in a Fe-Cr lattice [14]. Therefore, alternative strengthening particles should exhibit similar properties, and in the frame of this work, we intend to verify this hypothesis.

Ion irradiation is a fast and precise methodology that safely simulates the effects of neutron radiation on materials [17–19]. This methodology provides achieving high damage levels in a short time and the possibility of precise temperature and damage rate control (contrary to neutron radiation). Moreover, specimens submitted to ion irradiation are characterized by no (or negligible) residual radioactivity, what beneficially acts for operators' safety and cost of samples' post-processing [20]. Although ion irradiation is advantageous in introducing radiation defects into the crystal, it exhibits also some limitations. One of the most critical issues is a penetration depth. Due to the high electronic energy loss, ions lose energy very quickly. Consequently, a non-uniform energy deposition profile – and related damage profile – is observed [19]. Hence, implementing the ion irradiation technique causes the generation of a strongly modified thin layer with thickness varying from tens of nm up to 100 μm, depending on the energy and type of ions. Thus, investigating ion-affected layers requires implementing dedicated, highly precise examination methods to grasp both mechanical and structural changes triggered by radiation defects.

Motivation of this work is to evaluate the structural and mechanical response of steel strengthened by alternative oxides in a radiation environment. Since alumina and zirconia exhibit high thermal stability manifested by low Gibbs energy [14] (Fig. 1), they have been chosen as alternative substitutes for yttria in ODS material. Radiation damaging of materials structure has been performed by Ar ion irradiation, which allowed to creation strongly affected layer. Investigation of ion-irradiated zone allows for the assessment of the behavior of alternative ODS steel and description of mechanisms occurring under radiation damage.

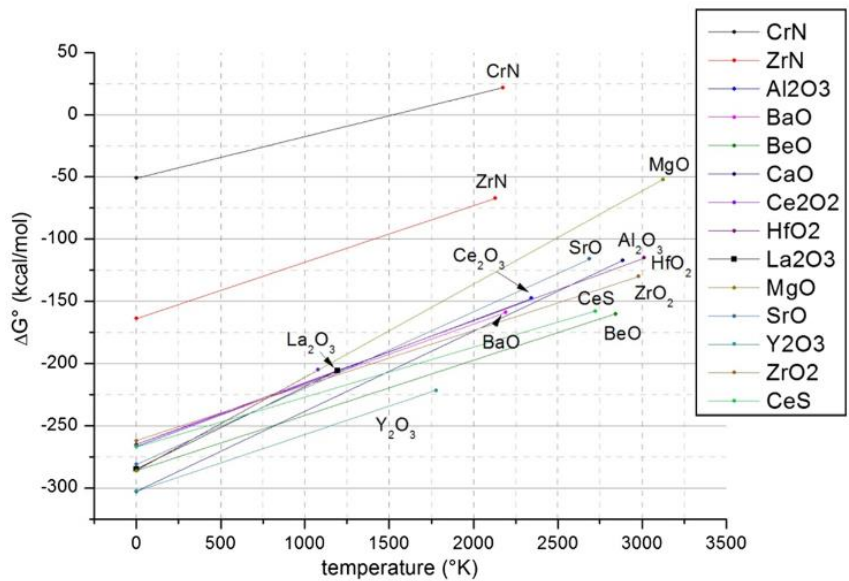

**Figure 1.** Gibbs-free-enthalpy of various oxides and nitrides [14].

## 2. Materials and methods

### 3.1. *Research materials*

Four materials have been fabricated using commercially available elemental powders provided by Alfa Aesar. The powders mixtures with nominal composition are presented in Tab. 1. Powders were mechanically alloyed (MA) in a high-energy planetary ball mill Retsch PM 100 with the rotation speed of 300 rpm in a high-purity Ar atmosphere for 50h. The ball-to-powder (BPR) ratio was set at 10:1.

**Table 1.** Nominal composition of manufactured research materials.

| Type of material | Cr [%wt.] | W [%wt.] | Ti [%wt.] | Refractory Oxide [%wt.] | Fe [%wt.] |
|---|---|---|---|---|---|
| Non-ODS | 12 | 2 | 0.3 | - | balance |
| Yttria-ODS | | | | 0.3 $Y_2O_3$ | |
| Zirconia-ODS | | | | 0.3 $ZrO_2$ | |
| Alumina-ODS | | | | 0.3 $Al_2O_3$ | |

In the next step, mechanically alloyed powders were consolidated in Φ25 mm graphite die using Spark Plasma Sintering (SPS) technique. The process was conducted at a temperature of 1100°C under a pressure of 50MPa. To prevent excessive grain growth and carbon diffusion (originating from the graphite die), consolidation was performed under a heating rate of 100°C/min. The materials holding time at maximum temperature was limited to 10 minutes. The whole process was conducted under a vacuum environment ($5 \times 10^{-5}$ mbar). As a result of the abovementioned procedures, four bars of bulk materials were obtained (Fig. 2). In order to obtain more homogenous materials, fabricated samples were heat-treated at 800°C in a high purity Ar atmosphere for 30min. After the normalization process, samples were cooled down in the air. Manufactured materials were sampled using wire electric discharge machining (WEDM) technology parallel to the bar axis (Fig. 2). WEDM cutting technique provides samples' surface free from residual stresses related to the tooling of material.

For further investigations, high-quality mirror-finished surfaces of the specimens were provided. As a first step, samples were submitted to the standard polishing procedure with sandpaper up to 1000x. The second step involved electro-polishing at a temperature of -10°C using a mixture of ethanol and perchloric acid.

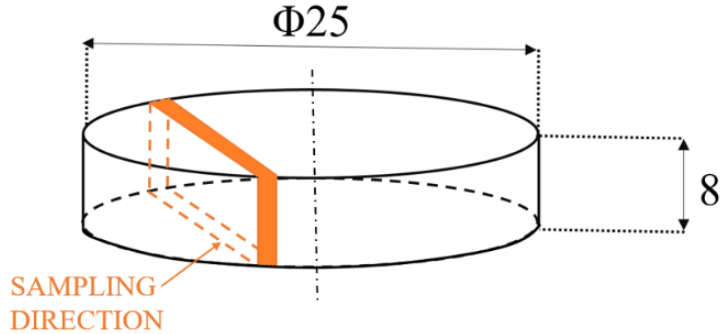

**Figure 2.** The geometry of manufactured samples with sampling manner. Dimensions in mm.

## 2.2. *Density measurements*

The density of fabricated materials was investigated by implementing the classical Archimedes principle using the Radwag analytical scale. As a medium, distilled water at room temperature was used. The weighting of each specimen was repeated at least five times, whereas samples between measurements were carefully dried for several hours in a dryer machine.

## 2.3. *Microstructure characterization (SEM/TEM)*

Structural observations were performed using a scanning electron microscope (SEM) Helios 5 UX provided by ThermoFisher Scientific. During observations, Electron Backscatter Diffraction (EBSD) and Energy Dispersive X-ray Spectroscopy (EDS) detectors were used. Electron transparent lamellas of pristine and ion-irradiated specimens were prepared by using the Focused Ion Beam (FIB) lift-out technique. Thin specimens were cut from previously prepared cross-sections manufactured for the above-mentioned SEM observations. During thinning of lamellas beam acceleration was gradually decreased in order to minimize the impact of gallium ions on observed microstructures. General observations of lift-out lamellas were conducted under scanning transmission electron microscopy (STEM) mode. Detailed analysis of irradiated areas was performed on a transmission electron microscope (TEM) JEOL JEM 1200EX II system operating at 120kV. For materials' comparison purposes, dislocation densities were estimated using line intercept method as presented in work [21]. Specimen thickness was determined with the use of convergent beam electron diffraction by measuring Kossel-Moellensted fringes spacing.

## 2.4. *Ion irradiation*

Rectangle shape samples with geometry 8 x 12 x 1 mm were submitted to ion irradiation, issuing Ar-ions with an energy of 500 keV. The damage profile (Fig. 3) was calculated using the Full Cascade mode with the SRIM program [22]. Specimens were irradiated with fluences $1 \times 10^{14}$ ions/cm², $1 \times 10^{15}$ ions/cm² and $5 \times 10^{15}$ ions/cm² what correspond to 0.1, 1.0 and 5.1 dpa, accordingly. Damage (dpa) calculation was performed according to the procedure described in the work of Agarwal *et al.* [23]. A thermocouple controlled the temperature of the samples during the irradiation process, and it did not exceed 70ºC. The temperature limitation during irradiation keeps migration rates of interstitials and vacancies lower and retards thermally-activated processes (such as dislocation climbing). Consequently, further rearrangement and interaction of mobile lattice defects leading to their reduction are limited [24]. As a result – the damage profile obtained using SRIM software may be regarded as unaltered due to the thermal effects during the irradiation process [25]. As a reference, an additional set of samples with masked surfaces were placed inside of ion accelerator facility.

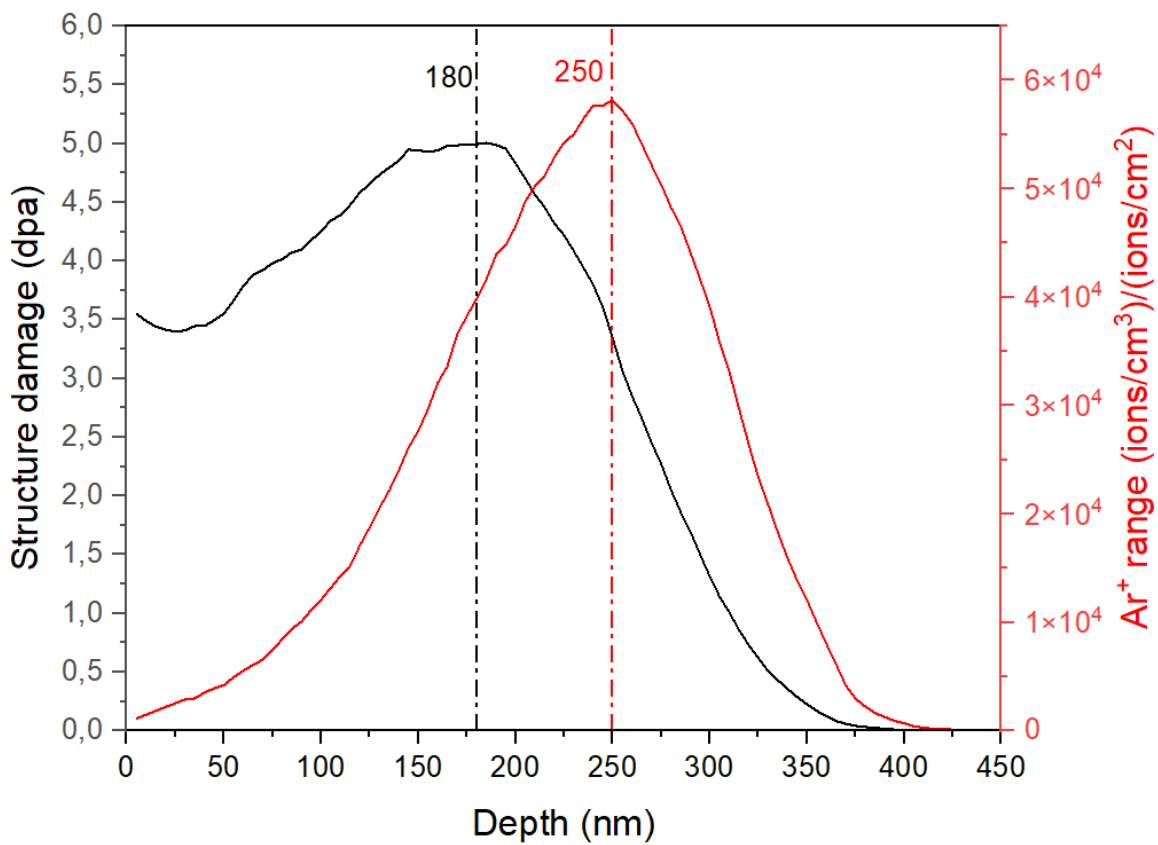

**Figure 3.** Damage accumulation (black axis) and irradiated Ar concentration (red axis) in manufactured ODS alloys as a function of depth calculated by SRIM with full-cascade mode [22]. The depth positions at 180 nm and 250 nm are marked with dashed lines separately as the maximum values of each profile.

## 2.5. *Grazing Incidence X-Ray Diffraction (GIXRD)*

Analysis of the changes in the ion-modified layer was conducted by implementing Grazing Incidence X-Ray Diffraction (GIXRD). Measurements were carried out using Bruker AXS (Advanced X-Ray Solutions) D8 Advance with a Cu X-ray source. The diffractometer scanning was performed from 20º to 145º of the 2θ angle with a constant ω angle of 1º. These parameters allowed for the collection of diffraction data which in 95% originated from the maximum depth of about 230 nm. The probed volume of the material corresponds to the depth at which the peak damage was recorded (Fig. 3). Analysis of the obtained datasets was conducted using DIFFRAC.TOPAS database.

## 2.6. *Nanoindentation*

Nanoindentation tests have been performed on a NanoTest Vantage System provided by Micro Materials Ltd. It has been well documented that the tip shape plays a significant role during low-load nanoindentation [26]. For this reason, before the tests, the Diamond Area Function (DAF) of the indenter tip was calculated for each indentation depth. Calibrations were performed using Fused Silica (FS) as a standard material with defined mechanical properties. Measurements were conducted at room temperature with a Berkovich-shaped diamond indenter (Synton-MDP). Indentations were performed in single force mode by using 11 forces in the range from 0,1 mN to 3 mN. Each measurement was repeated at least 30 times with 10 μm spacing between the indents. Loading and unloading times for each experiment were set for 5s and 3s, respectively, while dwell time was set for 2s. Nanomechanical properties were extracted from the nanoindentation load-displacement curves (L-D) by implementing well known Oliver and Pharr method [27]. Due to the presence of a plastic zone developed under the indenter tip, the penetration depth of the indenter during the experiment should not exceed 10% of the thickness of investigated layer [26]. Thus, the estimated depth of specimens' response during described experiments is within 20-200nm and includes the interest area – peak damage (Fig. 3).

One should explain that the nanoindentation technique provides information from the limited volume of the material (such as a thin layer). By probing such a small amount of the material, one should consider that every surface imperfection and local defect strongly influences obtained results. However, the precision of this methodology and reliability of acquired mechanical properties (by comparison with higher scale tests) have been proved [3].

## 3. Results

### 3.1. *Density measurements*

Figure 4 shows obtained densities of the manufactured specimens. The gray line represents the value of the representative material's density – pure iron with a density of 7.87 g/cm$^3$ [28]. The presented data indicates that all materials are characterized by a density of at least 98% of the reference value. These results suggest that used consolidation process parameters were correct and allowed to manufacture high-density bulk material with a negligible volume of pores and/or bubbles.

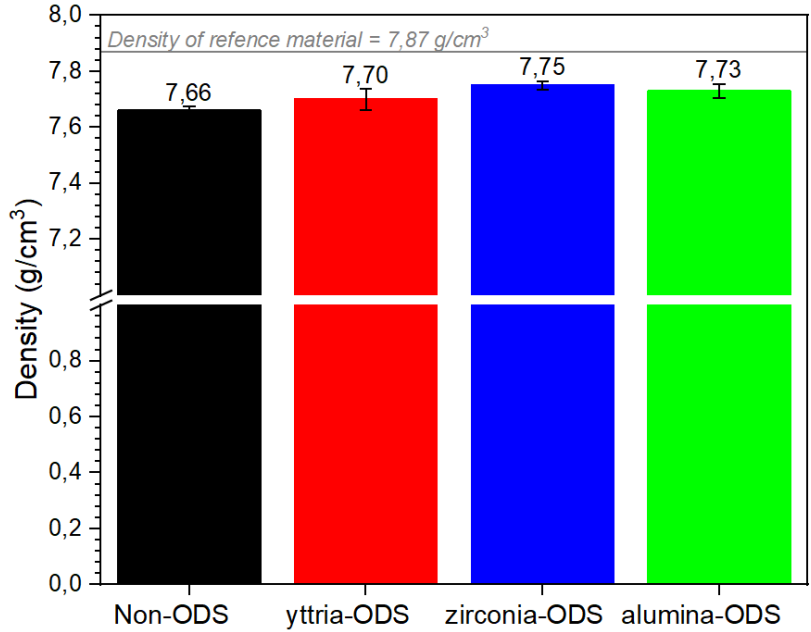

.

**Figure 4.** The density of manufactured specimens. The density of the reference material – pure iron [28] - is marked with a gray line.

### 3.2. *Microstructure characterization (SEM/TEM)*

Results of surface observations are presented in Figure 5. General analysis of obtained SEM images clearly shows that implemented consolidation technique allowed to obtain dense materials without visible pores or voids. A more detailed investigation revealed significant structural differences between fabricated materials. A net precipitates located in the proximity of the grain boundaries is well visible in non-ODS steel (Fig. 5A). EDS analysis of these precipitates revealed that they are enriched in chromium, tungsten, and iron. A similar effect has been registered in alumina-ODS steel samples. However, in this case, the precipitates were observed both in grain boundaries and in the grains' volume (Fig. 5D). Observations of yttria-ODS (Fig. 5B) and zirconia-ODS (Fig. 5C) samples revealed non-preferential arrangement of particles and their location seems to be random.

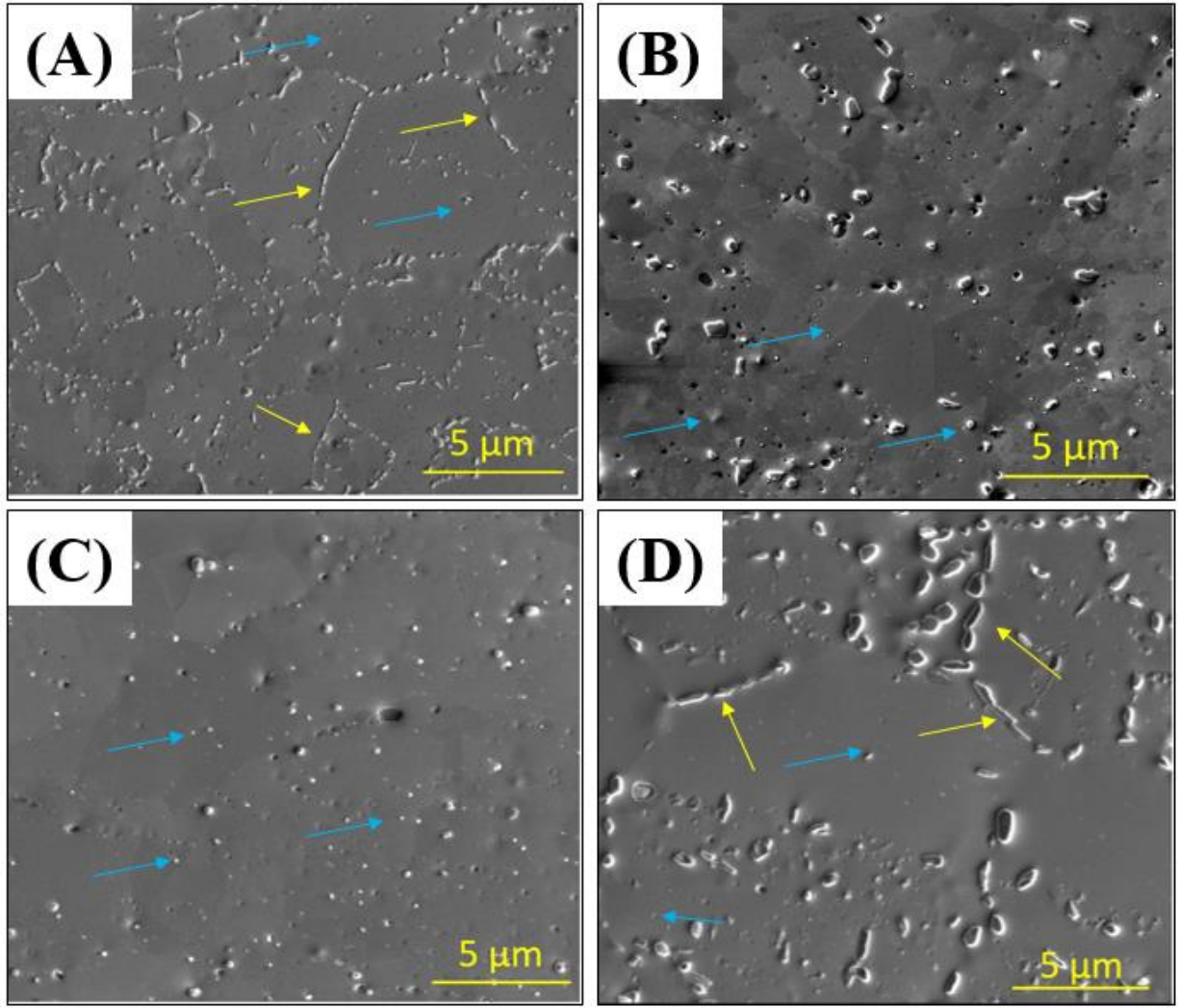

**Figure 5.** SEM images of manufactured research materials: (A) non-ODS (B) yttria-ODS (C) zirconia-ODS and (D) alumina-ODS. Yellow arrows indicate precipitates in grain boundaries, whereas blue arrows point to particles localized trans-granularly.

Detailed analysis of lamellas extracted from ODS steels strengthened with alternative oxides (alumina or zirconia) has been performed using STEM and TEM techniques. Recorded STEM images (Fig. 6 and Fig. 7) confirmed presence in both alternative ODS materials, strengthening particles localized inter- and trans-granularly. However, a larger number of particles in alumina-ODS specimens are detected in close proximity to the grain boundaries. Furthermore, some inhomogeneity in the strengthening particles distribution may be noticed in this material - marked with a blue rectangle area with large clusters of particles (Fig. 7). Moreover, two populations of particles (fine and coarse) are observed. It must also be highlighted that strengthening precipitates investigated in alumina-ODS are much larger than particles found in zirconia-ODS specimens. Chemical examination by EDS technique of these particles has revealed their enrichment in oxide, chromium, and titanium elements.

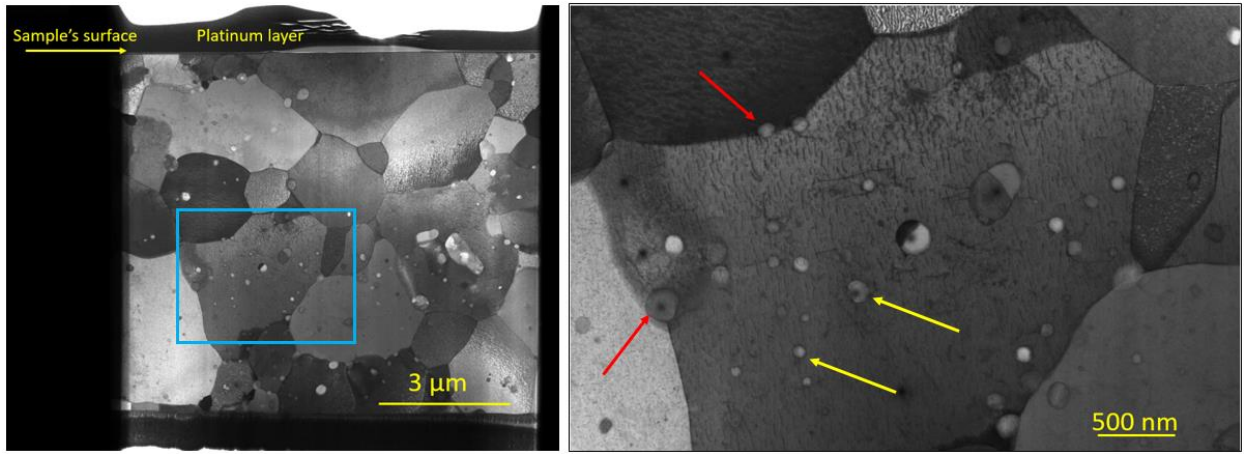

**Figure 6.** STEM observations of zirconia-ODS specimens. The blue rectangle corresponds to the figure's area on the right side. Red arrows indicate precipitates in grain boundaries, whereas yellow arrows point to particles localized intra-granularly.

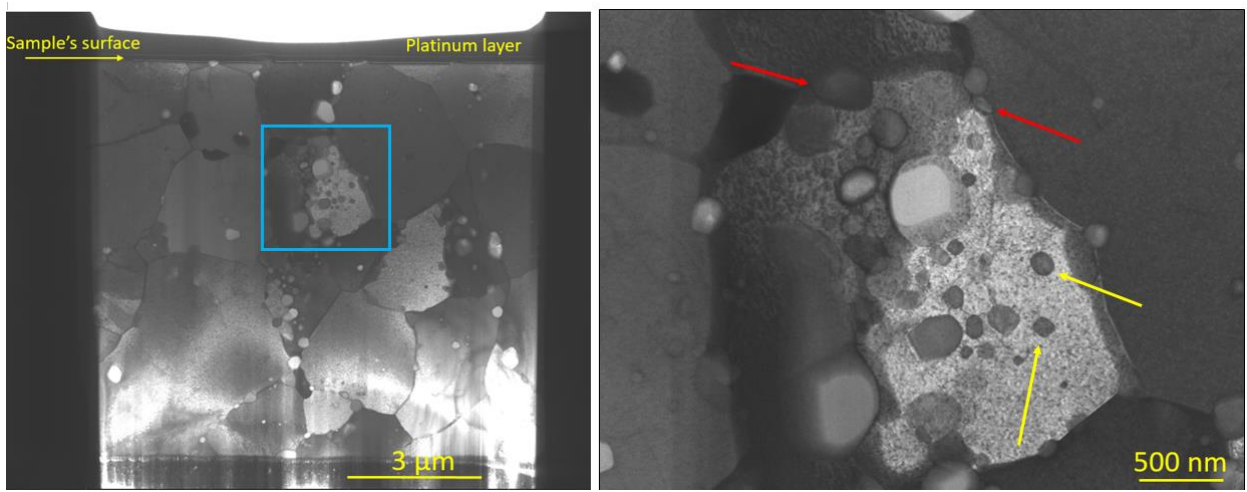

**Figure 7**. STEM observations of alumina-ODS specimens. The blue rectangle corresponds to the figure's area on the right side. Red arrows indicate trans-granular precipitates, while yellow arrows point to intra-granular particles.

The subsequent structural analysis was focused on grain size and their orientation. EBSD study confirmed that produced materials have BCC structure with random orientation of equiaxed grains (Fig. 8). A bimodal distribution of grains characterizes every manufactured material. Both fine and coarse grains are observed in all studied specimens. The estimated size of fine grains in zirconia-ODS is in the range of 0.7 – 1.4 µm, whereas in alumina-ODS specimens - in the range of 0.7-0.9 µm. Size of coarse grains has been estimated for 2-6 µm in both alternative oxide-ODS materials. However, numerous larger grains may be noticed in these materials. Furthermore, contribution of coarse grains in alumina-ODS steel is stronger than in zirconia-ODS. The largest grains (approx. 0.9-2.5 µm) were observed for non-ODS steel, while the size of the prevailing majority of the grains exceeds 1.5 µm in diameter. Numerous grains with estimated diameter above 5 µm are also well visible (Fig. 8A). Yttria-ODS steel exhibits the finest grain structure, where the majority of grains are lower than 0.5 µm. Moreover, contribution of coarse grains in yttria-ODS steel is much lower than in specimens strengthened by alternative oxides.

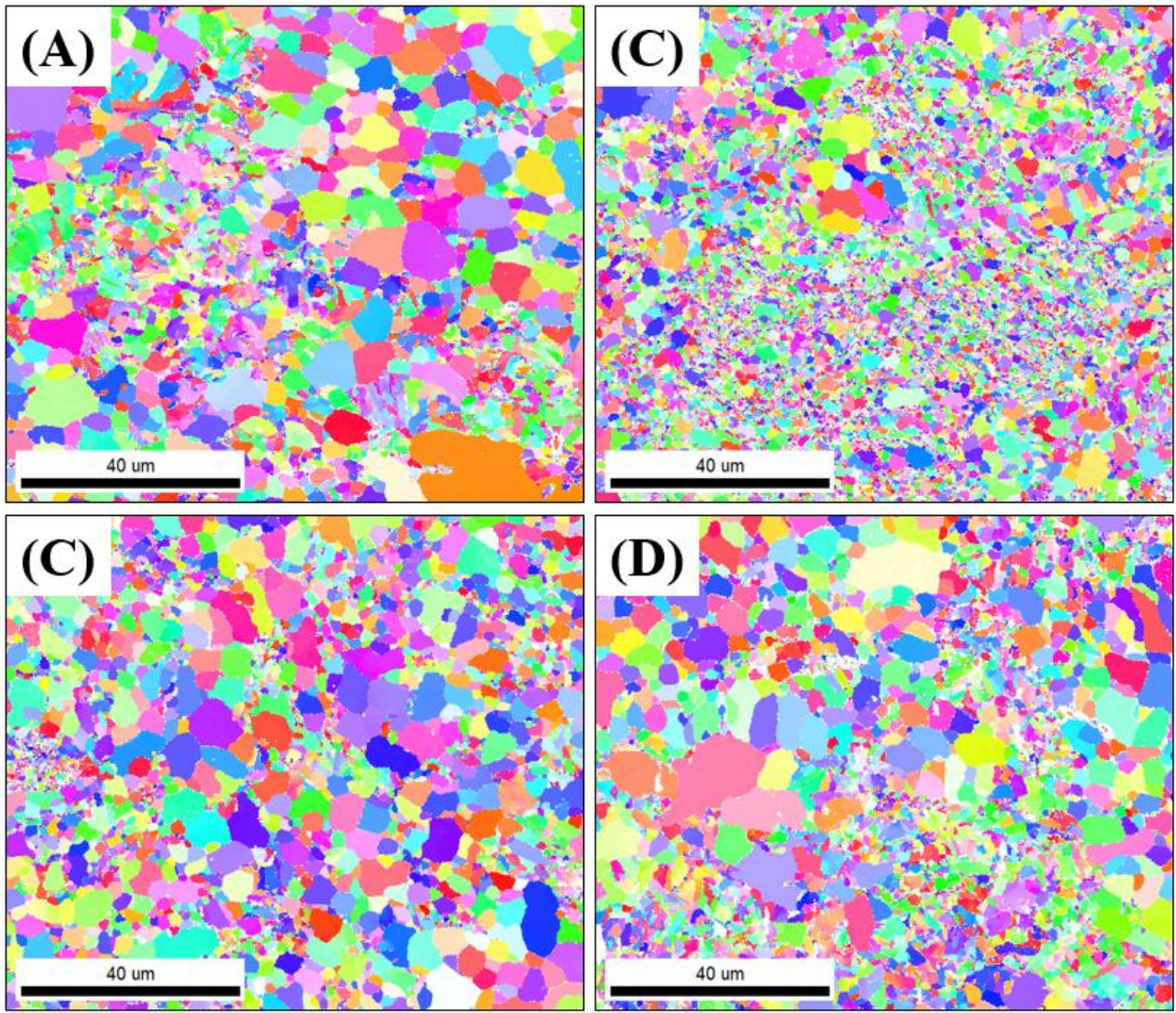

**Figure 8.** EBSD analysis of manufactured materials: (A) non-ODS (B) yttria-ODS (C) zirconia-ODS and (D) alumina-ODS.

The structural investigation by TEM revealed the presence of many defects in the irradiated layer (Fig. 9). One can see that the density of black dots (most probably small dislocation loops or stacking fault tetrahedra – SFT [29]) and dislocations (black lines) is continuously increasing with increasing depth from the samples' surface. This phenomenon continues to increase until maximum damage (the dark area) reaches a depth of about 220 - 230 nm. Moreover, the estimated density of defects developed in the Ar-modified layer has been done with a 50 nm step. This has been correlated with the SRIM profile and is presented in Fig. 10 (marked as blue dots). One may note that the tendency change of calculated defect density closely matches the estimated SRIM damage profile. This is particularly visible for the alumina-ODS specimen, where a high-level correlation between experimental data and the SRIM model has been found. However, TEM observations of ODS steel strengthened with alternative oxides revealed that peak damage is localized slightly deeper (Fig. 9) than estimated by using SRIM software. As predicted by the SRIM code, peak damage should be localized below 200nm, whereas TEM analysis revealed its presence approx. 30 - 40nm deeper. It must be highlighted that this difference is not meaningful and is related to the differences between SRIM code assumptions (calculations made for ideal alloy) and real experiments (some fluctuation of ions' energy as well as chemical composition and its local inhomogeneity). Depending of the type of the ions and chemical composition of the target, the thickness of irradiated layer may differ from SRIM predictions from 15 even up to 50% [30]. Moreover, it is well known that the radiation damage profile strongly depends on the metal structure. In the literature, it is called the "long-range effect" [29,31]. This phenomenon is related to the processes taking place during irradiation. Extremely high temperatures occurring close to the ions track and lasting for a very short period (thermal spike) generate strong stress field gradients. This may lead to the multiplication of dislocations from pre-existing loops. Such defects propagate into the bulk escaping from the cascade area. It is well known that Peierls force opposing the dislocation glide through the structure is lower in FCC structure (where atom density is higher). Thus, FCC-structured materials are characterized by a damage profile usually greater than a predicted model. The opposite situation occurs in BCC materials – in this case, shallower damage profiles are recorded [29,31,32]. In the discussed experiment, numerous pre-existing dislocations and vacancies are present. A high density of defects leads to a strong intensification of the dislocations' multiplication during the thermal spike. Furthermore, it must be highlighted that during a thermal spike, the mobility of the vacancies is also intensified. The number of vacancies during irradiation is increasing, which also affects mobility of the dislocations. Thus, it seems that a large number of pre-existing defect affects the profile damage, leading to its' broadening instead of shallowing (as discussed in the literature). This phenomenon suggests that defects generated as a result of irradiation interact with pre-existing defects, distorting the material's true nature. Finally, one must remember that we are considering very low (in comparison to the literature) irradiation energy (shallow irradiation depth), and the observed difference is very small; hence observed difference in profile depth is within the measurement error.

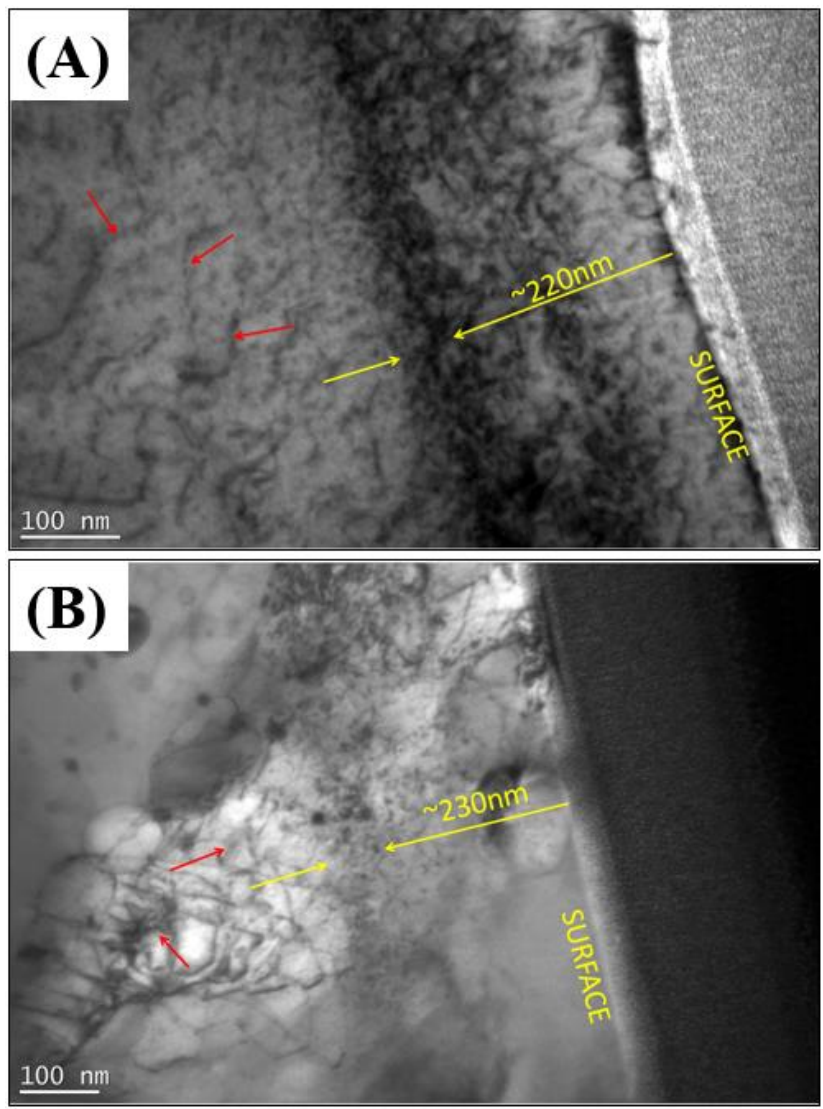

**Figure 9.** Results of TEM observations of materials submitted to ion irradiation with dose up to $5\times10^{15}$ Ar$^+$ ions/cm$^2$: (A) zirconia-ODS and (B) alumina-ODS. Yellow arrows indicate the dislocations at the peak damage, and the red arrows point to defects located below the irradiation-affected zone.

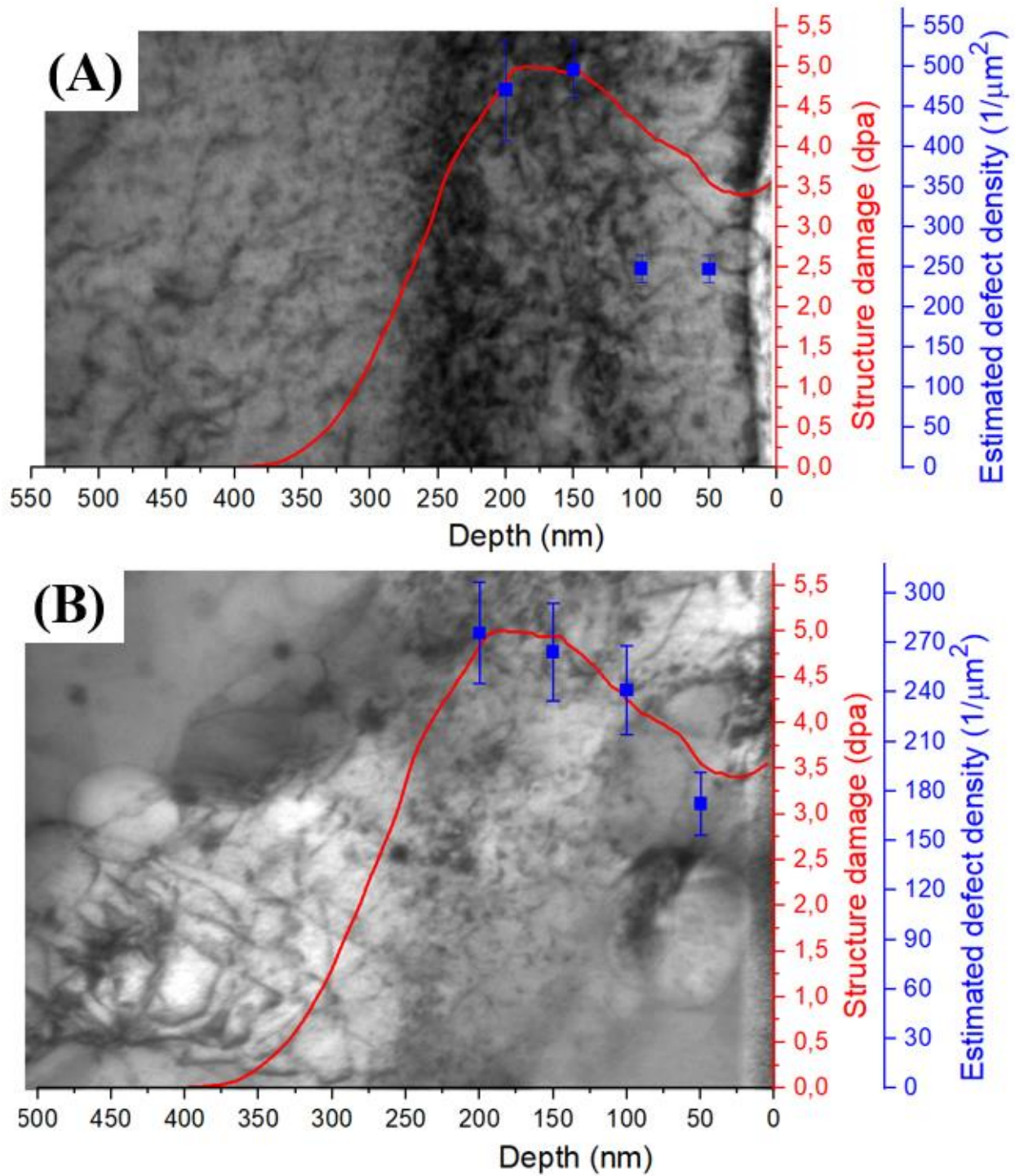

**Figure 10.** Calculated the SRIM damage profile and estimated defect density with corresponding TEM images (see Fig. 9) of the studied materials: (A) zirconia-ODS, and (B) alumina-ODS.

Further examination of the specimens shows the presence of numerous structural defects located below the irradiated layer (Fig. 9). Appearance of these pre-existing defects (in the majority – dislocation lines, dislocation loops, and clusters of point defects) is related to the materials manufacturing history (mechanical alloying). It is well known that mechanical alloying includes heavy deformation of particles, which leads to introducing to the structure variety of crystal defects, such as vacancies, dislocations, stacking faults, and increased grain boundaries volume [33]. In addition, the high-temperature consolidation process and heat treatment are not sufficient to allow for the annihilation of these defects. Thus, numerous pre-existing defects are well visible. The estimated density of defects observed in investigated specimens is presented in Tab. 2. Dislocation density increases after ion irradiation for both types of materials. This is related to well-known radiation damage of the structure and generation of numerous crystal defects [34]. However, one can see that the dislocation density increment is much higher for zirconia-ODS steel than for alumina-ODS material. The estimated dislocation density in zirconia-ODS steel is more than twice of alumina-ODS specimens. Furthermore, a comparison of a number of pre-existing defects reveals that material strengthened by zirconia is characterized by a much lower value than alumina-ODS alloy. The response of zirconia-ODS steel to ion-irradiation is much more intensive than material with alumina. It must be highlighted that presented in this publication values of density of dislocation are only estimations calculated on a very limited area (several cubic microns of the lamella).

**Table 2.** Estimated density of dislocation in lift-out lamellas of ODS steels strengthened by alternative oxides. Calculations performed from Ar-affected zone (described as "Ion-irradiated") and below the modified layer (described as "Virgin").

| Material | Dislocation density $[1/{\mu m^2}]$ | |
|---|---|---|
| | Virgin | Ion-irradiated |
| **Zirconia-ODS** | 350 | 480 |
| **Alumina-ODS** | 104 | 215 |

### *3.3. Grazing Incidence X-Ray Diffraction (GIXRD)*

Alteration of structural features due to ion interaction with the matter has been examined employing GIXRD (Fig. 11). Experiments were performed on pristine specimens and submitted to Ar-ion irradiation with the highest fluence ($5 \times 10^{15}$ ions/cm$^2$). Due to chemical composition of research materials (see Tab. 1), predominant impact on diffraction patterns has two elements: iron and chromium.

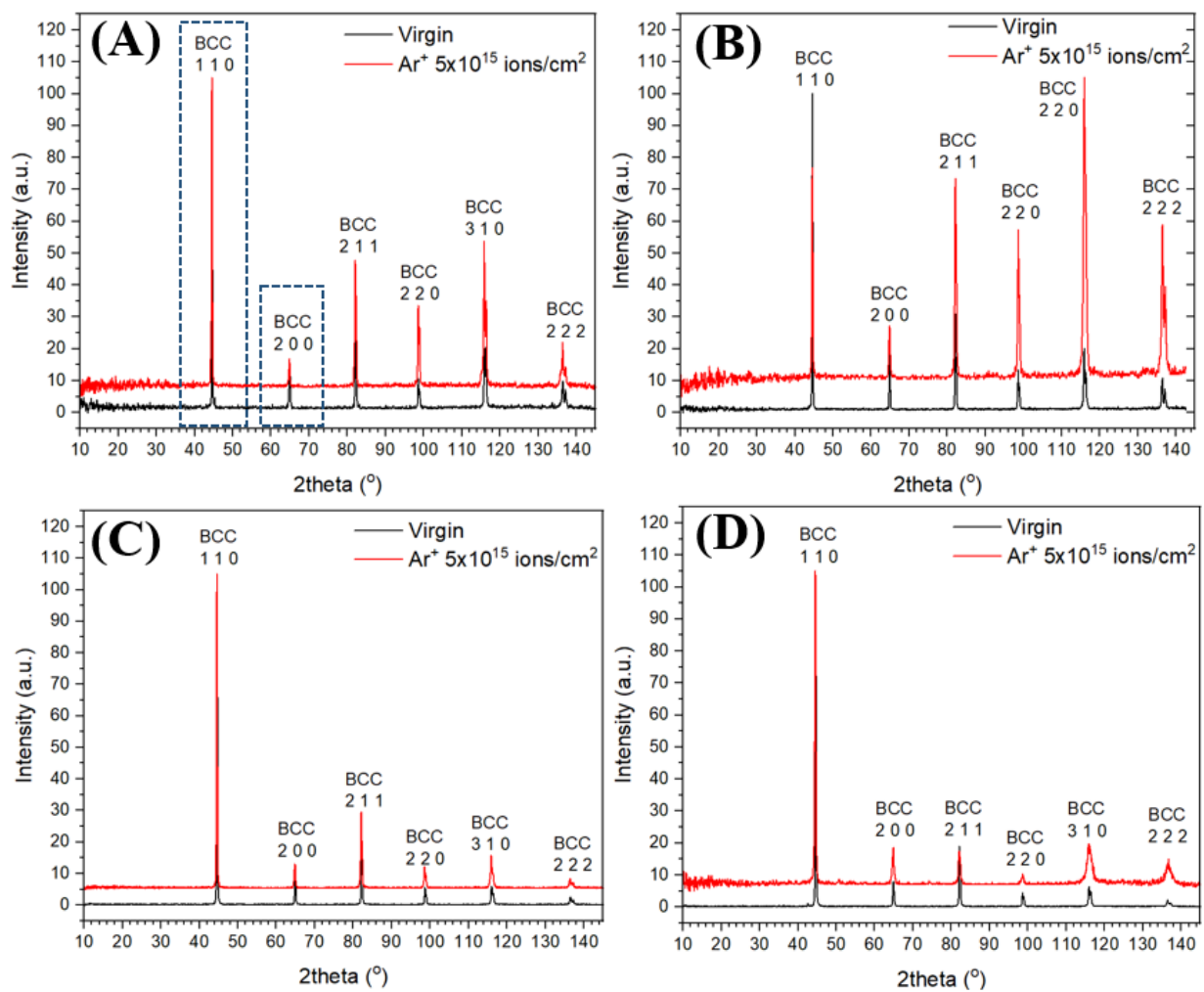

**Figure 11.** Grazing Incidence XRD diffraction patterns registered for: (A) non-ODS (B) yttria-ODS (C) zirconia-ODS, and (D) alumina-ODS. Areas marked in blue rectangles are magnified in Fig. 11 (peak 110) and Fig. 12 (peak 200).

In each material, several characteristic peaks for αFe-Cr solid solution (BCC structure) were registered (Fig. 11). Evaluation of recorded diffraction patterns has shown slight shifts of the peaks and their broadening. In order to obtain a higher clarity of the shifts mentioned above, Fig. 12 and Fig. 13 present the magnification of characteristic peaks. Due to a large number of registered signals, this publication presents displacements of only two of them (as an example) ‒ registered at $2\theta \approx 44.67°$ (which corresponds to 110 plane) and registered at $2\theta \approx 65.02°$ (which corresponds to 200 plane).

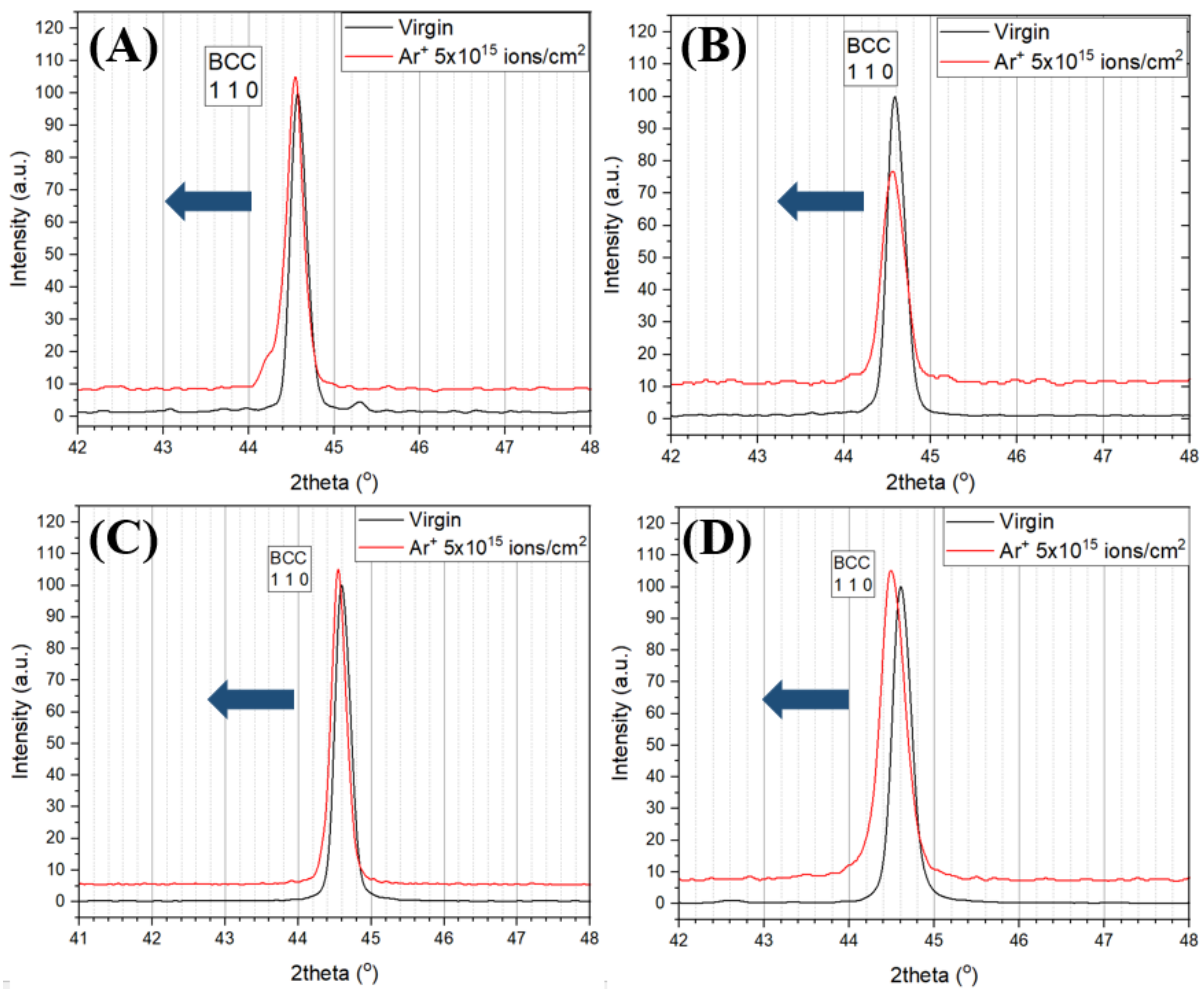

**Figure 12.** Magnification of peaks corresponding to 110 planes of αFe-Cr for ODS samples as received and submitted to Ar$^+$ irradiation up to $5\times10^{15}$ ions/cm$^2$: (A) non-ODS (B) yttria-ODS (C) zirconia ODS, and (D) alumina-ODS. Blue arrows indicate the tendency to alter after the 2θ position of recorded peaks.

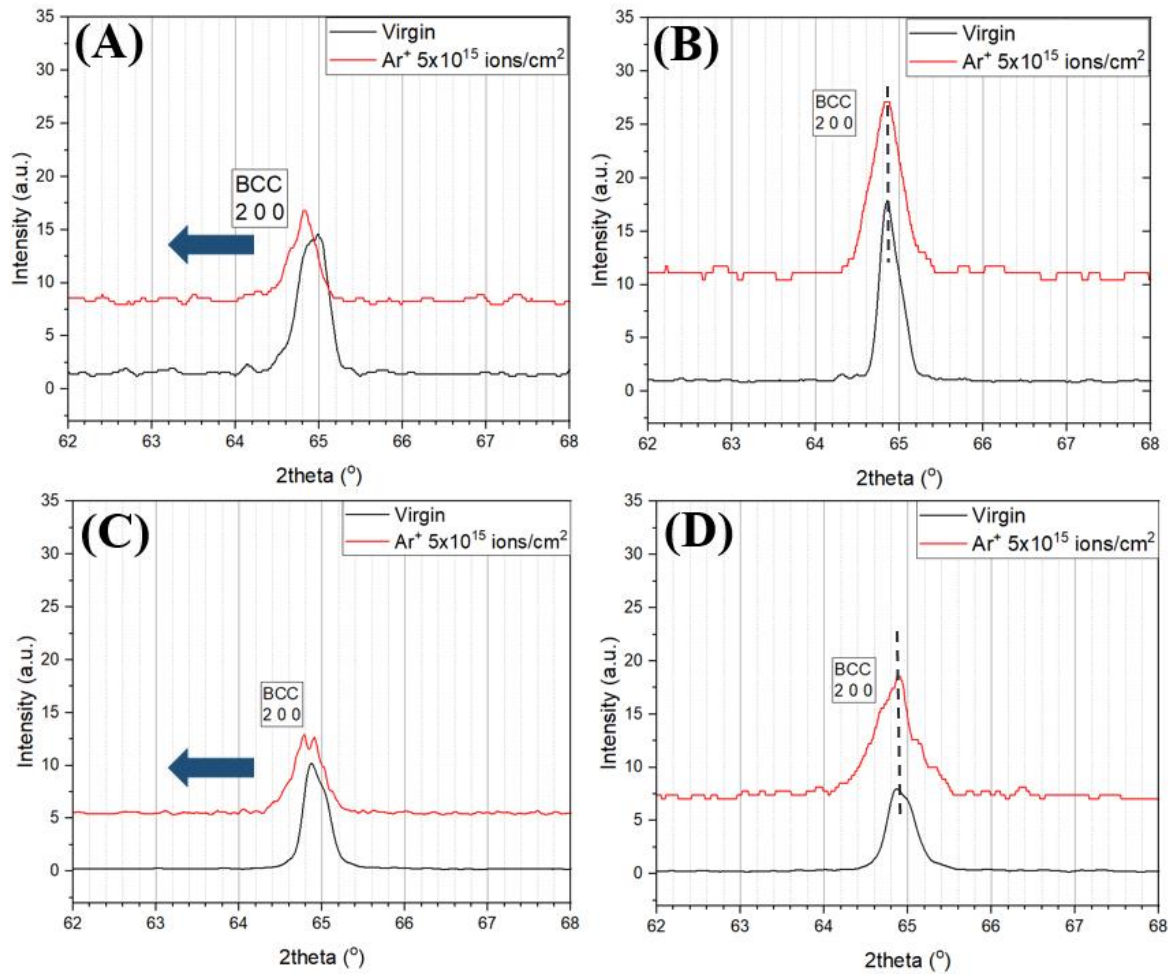

**Figure 13.** Magnification of the peaks corresponding to 200 plane of αFe-Cr for ODS samples as received and submitted to Ar-ion irradiation up to $5 \times 10^{15}$ ions/cm$^2$: (A) non-ODS, (B) yttria-ODS, (C) zirconia ODS, and (D) alumina-ODS. Blue arrows indicate the shift tendency of recorded peaks, while a dotted blue lines indicate their non-significant change of 2θ position.

One may notice that for the (110) plane a slight shift of the peaks into lower 2θ values is observed. This effect has been observed for all studied materials (Fig. 12). Shift of the (200) plane is inconsistent. The broadening of the diffraction peaks with no significant change in their position is recorded (Fig. 13). It is known that modifications of the diffraction peak shape and position is related to alteration of the crystal structure. Increasing inter-plane distance *d* is manifested by decreasing 2θ value [35], whereas broadening of the peak is related to lowering the crystals' size through the Debye Scherrer equation [36]. Detailed investigation of peaks' shifts and their broadening allows for the calculation of lattice parameters and microstrain, as presented in work [37]. Results of these calculations are shown in Tab. 3. One may notice that ion irradiation leads to increment in both lattice parameters and microstrains. The lowest structural changes are observed for yttria-ODS specimens, while non-ODS materials and zirconia-ODS steel exhibit similar behavior. The most significant difference has been recorded for alumina-ODS. Moreover, this material also displays the strongest increase of microstrain related to the ion – irradiation. In this case increase by one order of magnitude is observed, whereas recorded changes in other materials are much lower. It must be highlighted that collected signals in the geometry mentioned above originate in 95% from the maximum depth of 230nm. However, the impact of the non-modified material (below the irradiated layer) cannot be neglected. Furthermore, one must remember that the studied materials are polycrystalline and are characterized by fine grains (Fig. 8). Implemented measurements' geometry and the above-mentioned materials' structural features may impact Bragg's condition. For some specific angles in the examined volume of the material, the criterion may be met. However, some domains do not reveal their actual influence on whole bulk material. Furthermore, the level of damage in the investigated layer is not constant (Fig. 3), whereas the presented results should be assumed as values averaged through 230 nm of materials' depth. Nevertheless, the tendency of the diffraction peak change as a result of ion irradiation for different materials is still well-visible.

**Table 3**. Structural parameters calculated from GIXRD datasets obtained for ODS steels in as received state and submitted to Ar-ion irradiation: (A) lattice parameter and (B) microstrains.

| Material | State of the sample | Lattice parameter [Å] | Microstrain $\varepsilon_0$ |
|---|---|---|---|
| Non-ODS | Virgin | 2.873366(10) | 0.000379(6) |
| | Ion- irradiated | 2.874816(10) | 0.000320(30) |
| Yttria-ODS | Virgin | 2.873222(10) | 0.000305(7) |
| | Ion- irradiated | 2.873511(8) | 0.000433(7) |
| Zirconia-ODS | Virgin | 2.872246(12) | 0.000365(4) |
| | Ion- irradiated | 2.874645(6) | 0.000546(4) |
| Alumina-ODS | Virgin | 2.872399(15) | 0.000461(7) |
| | Ion- irradiated | 2.877135(17) | 0.001349(15) |

### *3.4. Nanoindentation*

Mechanical properties of the studied materials submitted to ion irradiation are presented in Fig. 14. Hardness (H) increase at a depth of about 40 nm is well visible for all manufactured materials. This effect is related directly to the peak damage generated by the ion irradiation process. As shown in work [38], due to the presence of a plastic zone located below the indent, the depth of signal's registration is much lower than the depth at which peak damage is located (Fig. 3). Recorded hardness increment as a result of radiation damage build-up is well-known, and in the literature this effect is called "irradiation hardening" [7,17,39,40]. However, the magnitude of this effect has never been estimated for steels strengthened by alternative particles. After reaching the maximum H value at about 40 nm depth, hardness gradually decreases with the increasing contact depth. This tendency is visible until the recorded H value slowly approaches the hardness of unmodified material (Fig. 13).

Figure 15 presents the hardness alteration of the studied materials in the function of increasing Ar-ion fluence. One may note that the more damaged by ion-irradiation structure is, the more intensive mechanical properties change is recorded. Yttria-ODS steel seems to be the most resistant to irradiation hardening effect. This material exhibits the lowest increase of the H value over the whole fluence spectrum. For the lowest fluences, yttria-ODS and zirconia-ODS exhibit similar behavior (increment of approx. 30%). The strongest effect (H increase), when comparing one fluence over all materials, has been recorded for non-ODS steel (increment above 40%). For higher fluences ($1 \times 10^{15}$ ions/cm$^2$), hardening by 40-45% of the initial value is observed for all ODS specimens. Non-ODS steel exhibits H increase by more than 50%. The hardness of the materials submitted to ion irradiation up to $5 \times 10^{15}$ ions/cm$^2$ increases at least by 70% for non-ODS, zirconia-ODS, and alumina-ODS, while for yttria-ODS' this parameter does not exceed the value of 60%.

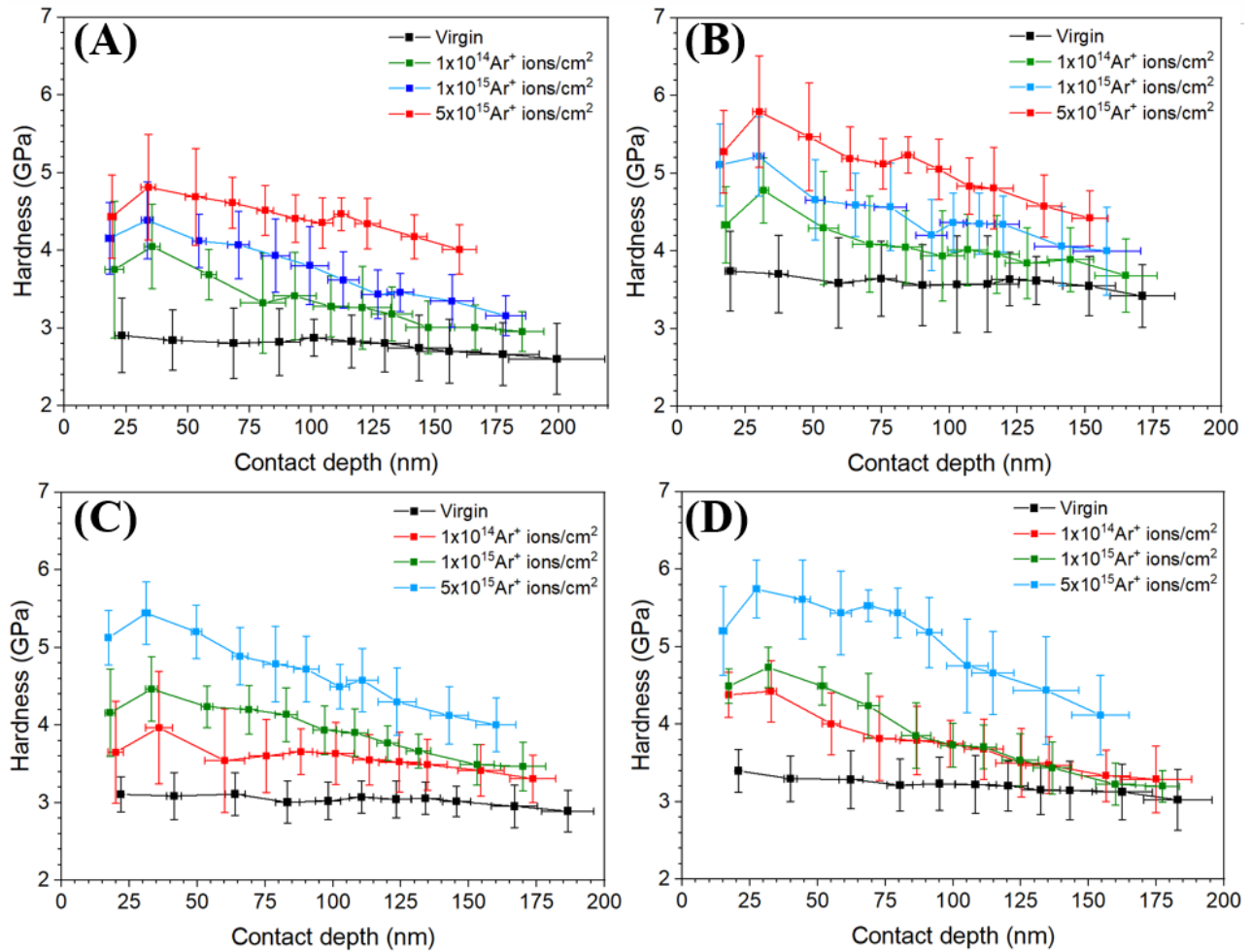

**Figure 14.** Hardness as a function of depth of all studied materials: (A) non-ODS, (B) yttria-ODS, (C) zirconia-ODS, and (D) alumina-ODS. Data were obtained by using the nanoindentation technique.

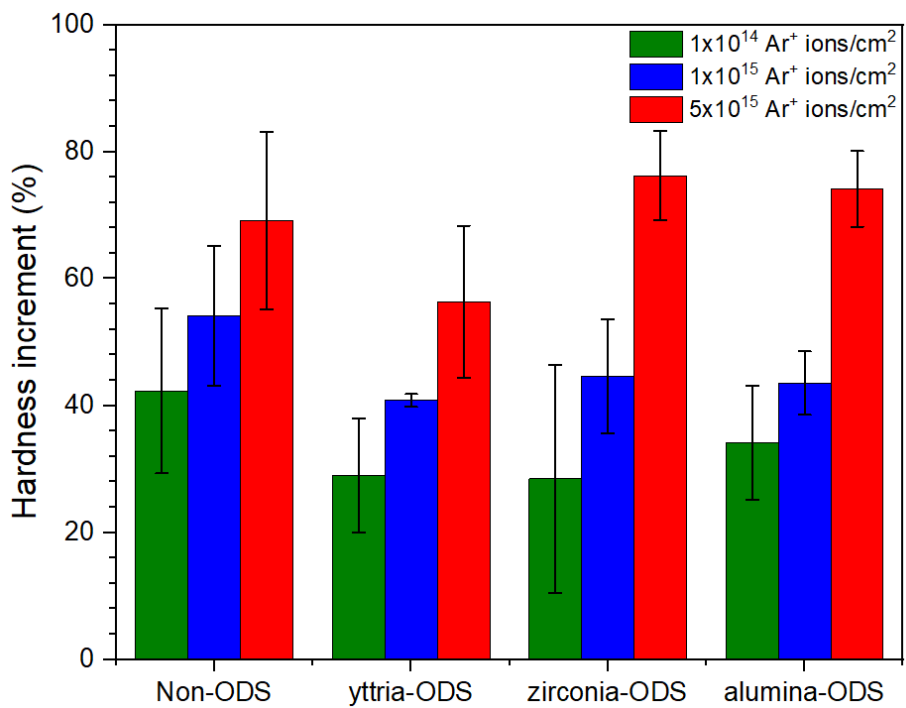

**Figure 15.** Calculated hardness increment measured for all manufactured specimens submitted to Ar$^+$ irradiation with different fluences.

## 4. Discussion

Structural analysis by SEM (Fig. 5) revealed the presence of precipitates in non-ODS reference material. These particles are enriched in chromium, tungsten, and iron and are inhomogeneously distributed over the grain boundaries. Reported findings suggest that observed precipitates may be identified as $M_{23}C_6$ carbides (where M is a mixture of Cr, W, and Fe). The same particles were reported in works [16,41]. The presence of carbon in manufactured materials is probably related to the consolidation process, where graphite dies were used. EDS analysis revealed that in alternative oxide-ODS specimens, large particles are enriched with chromium, oxide, and titanium. This data suggests that similar processes as observed in non-ODS samples occur, while instead of complex carbides – complex oxides precipitates are being found. Low enrichment with Al and Zr may be explained by their possible partial dissolution in the ferritic matrix. It seems that strengthening refractory oxide (zirconia or alumina) undergoes incomplete decomposition during the production process, and above-mentioned elements have been replaced by Cr atoms in complex oxides particles. The strong reaction between yttria and the lattice is well described in the literature. Several effects including: (i) refinement of strengthening particles [41], (ii) increment of their complexity (i.e., transformation in pyrochlore $Y_2Ti_2O_7$ [8], $Y_2TiO_5$ in hexagonal structure [9] or orthorhombic [42]) as well as (iii) formation of an interface between the particles and the matrix in yttria-ODS steels [43] have been reported. Performed investigation of alternative oxide ODS materials suggests the existence of similar interaction between matrix elements and refractory oxides (zirconia or alumina). However, these effects remain unclear, and they require further investigation.

Conducted GIXRD tests on as-received and submitted to ion-irradiation specimens revealed slight diffraction peak shifts and their broadening (Fig. 11 - 13). Displacement of the peaks from their original positions and their shape fluctuation may have various reasons. Broadening of the diffraction signal occurs when there is a loss of crystallinity [35]. Hence, the more crystal defects are created, the broader the peak becomes. Furthermore, the deviation of the diffraction peak from its 2θ position is strictly related to the change in the inter-planar distance $d$ [35]. Both effects (peak shifts and broadening) may occur independently, although one effect may intensify the other. Introduction to the ideal crystal structure radiation defects such as dislocations, stacking faults, or point defects leads to loss of crystallinity of the metal. It also generates stresses in the structure which may have long- or local- character. Reported effects may impact inter-plane distance $d$. Depending on the situation, an increase or decrease of the $d$ value demonstrates that tensile or compressive stresses are realized in the structure. Performed GIXRD investigation revealed that lattice parameter , as well as microstrains in all materials, increase due to ion-irradiation (Tab. 3). It is noteworthy that increment of lattice parameter and microstrain value is much less intensive for zirconia-ODS specimens than for alumina-ODS. This observation suggests that the structure of zirconia-ODS steel seems to be more resistant to radiation damage. To fully understand this observation, analysis of EBSD observations (Fig. 8) and calculated each specimen's average grain size should be included. It must be highlighted that the presence of grain boundaries plays a crucial role in material behavior under irradiation conditions. Bai *et al.* [44] proposed effective annealing phenomena, where grain boundaries act as defect sinks for their own interstitials (created both as primary knock-out atoms (PKA) or in cascade effects [24]). Once interstitials are loaded to the grain boundary, they are emitted to the matrix to recombine with produced vacancies. This process is energetically favorable to annihilating immobile radiation vacancies trapped in the bulk material by interstitials created by radiation damage. His model assumes that described diffusion process proceeds at a close distance from the grain boundaries (up to 10 Å). Thus, the decrease of grain boundary volume also leads to the limitation of the phenomena mentioned above. Furthermore, ion irradiation campaigns have been performed at room temperature, where the mobility of vacancies in the structure is very low [24]. Therefore, the proposed annealing phenomena is limited to point defects generated nearby grain boundaries. Excess radiation-induced point defects generated at a higher distance from grain boundaries is utilized in spontaneous local recombination processes if the distance between vacancy and interstitial is within several lattice periods [24]. Since zirconia-ODS steel is characterized by a larger volume of grain boundaries than alumina-ODS, it may be assumed that radiation-induced defects annealing is more effective in this material. Consequently, the calculated increase of the lattice parameter value and the level of microstrains are much less intensive for zirconia ODS than for alumina ODS. Moreover, the presence of Ar atoms trapped in examined irradiation affected zone cannot be neglected (Fig. 3). Nevertheless, this factor may be assumed as a constant for all materials, because specimens were submitted to the same level of ion radiation.

It has been proved that during radiation damage strengthening particles act as defect sinks [40–42]. As Bai *et al.* [44] proposed, similar phenomena may be applied to the interaction between defects and strengthening particles [45]. Hence, the interaction between sinks (such as strengthening particles) and radiation-induced defects also significantly influence materials behavior under irradiation. According to SEM observations (Fig. 5-7), particles in zirconia ODS steel may be characterized as fine precipitates randomly distributed in a matrix. Strengthening particles in alumina-ODS specimens are much larger and localized mainly in the grain boundaries or their proximity. Since the mobility of vacancies at room temperature is strongly limited [24], homogeneous distribution of defect sinks (such as strengthening particles as well as grain boundaries) is the factor that efficiently retards radiation-induced degradation. It provides a large area for the recombination of radiation defects. For this reason, the structure of zirconia-ODS specimens (large volume of grain boundaries and fine precipitates homogenously distributed in the matrix) seems to be much more resistant to radiation damage than the structure of alumina-ODS. These features also have a relevant influence on recorded differences between both materials during GIXRD analysis.

Nevertheless, neither alternative oxide-ODS steel exhibits a better response for ion irradiation than reference yttria-ODS material. Although all introduced oxides (yttria, zirconia, and alumina) may be characterized as refractory [46], the investigation revealed diametrically different structures and radiation resistance. The shape of the precipitates and their distribution suggest that chemical elements of zirconia and alumina may partially dissolve in the matrix, whereby the most noticeable precipitates may be classified as $M_{23}C_6$ carbides [16,47]. It is known that incoherent particles (as yttria in ODS steel) act as neutral defect sinks, where are no preference for capturing a specific type of defect (vacancy or interstitial) exist [24]. Capturing defects in incoherent particles is proportional to the diffusion coefficient of the defect and its gradient between the matrix and the surface of the sink. Noticeable carbides - coherent or semi-coherent particles - act as variable biased sinks with a limited capacity [24]. These differences in coherency and related capacity for radiation defects recombination underlay in fundamentals of observed dissimilarities between studied specimens. Yttria-ODS specimens exhibit a less intensive response to ion irradiation. This is demonstrated by the lowest fluctuation of the *d* parameter, an increase of the structural microstrains (Tab. 3) as well as lowest hardening effect (Fig. 15). Therefore, it seems that yttria not only sufficiently captures radiation point defects, but also efficiently limits grain coarsening (Fig. 8). As mentioned above – grain boundaries are considered as areas with intensive self-annealing properties. According to obtained results, zirconia and alumina probably partially decompose during the production process. This behavior does not provide a fine-grained structure. However, according to the presented structural results, one may assume that introducing zirconia to the ferritic matrix has a more beneficial effect on irradiation resistance than introducing alumina particles. Although, it must be highlighted that, as suggested in this work, partial decomposition of strengthening alternative oxides must be confirmed by further examination.

Conducted nanoindentation tests (Fig. 14) show that non-ODS reference steel displays the lowest mechanical parameters in the as-received state. One can also notice that this material is the most vulnerable to radiation hardening (Fig. 15). Since non-ODS steel is not strengthened with any refractory oxides, and is characterized by the largest grain size, typical strengthening mechanisms (by grain boundaries and dispersion strengthening) are not efficient [9]. According to the mechanism of the self-annealing process during radiation damage, obtained results are not surprising. Analysis of all ODS steel shows that yttria-ODS steel displays the highest radiation resistance, which is demonstrated by the lowest hardening effect (Fig. 15). It must be highlighted that the behavior of all ODS-material exposed to ion irradiation up to $1 \times 10^{15}$ ions/cm$^2$ fluence is very similar. The hardening of materials in this environment is on the same level, independently of what type of strengthening oxide has been introduced to the matrix. Increasing the fluence up to $5 \times 10^{15}$ ions/cm$^2$ leads to the differentiation between research materials. Hardening of ODS materials strengthened by alternative oxide submitted to irradiation up to $5 \times 10^{15}$ ions/cm$^2$ is at the same level as hardening of non-ODS steel. These results suggest that the proposed self-annealing process in zirconia- and alumina-ODS specimens have the same efficiency as in the yttria-ODS steel for doses up to $1 \times 10^{15}$ ions/cm$^2$. Consequently, it indicates that the rate of radiation defects recombination in grain boundaries and strengthening particles and the rate of defect generation in all ODS-steel is similar, irrespectively from the type of introduced particles. Hence, alternative strengthening oxides provide similar radiation resistance as reference yttria particles up to the dose $1 \times 10^{15}$ ions/cm$^2$. Thus, it may be concluded that the self-annealing process below the described radiation level maintains the same efficiency regardless of the coherency of strengthening particles. Above this value, structural differences between materials such as grain boundary volume and strengthening particles seem to play a crucial role in retarding radiation damage. Conducted examination reveals that at higher radiation doses, the capacity of structural elements for radiation defects annihilation is much higher for yttria-ODS than materials strengthened by alternative oxides.

Obtaining the TEM images (Fig. 9) allowed for estimation of dislocation density change due to ion irradiation (Tab. 2). One may notice that increment of dislocation density is much more intensive for zirconia-ODS than for alumina-ODS. Moreover, the level of pre-existing dislocation in bulk material varies by order of magnitude depending on the type of strengthening oxide introduced to the matrix. In zirconia-ODS specimens, dislocations seem to have a form of short, spontaneously arranged lines. In contrast, in alumina-ODS specimens, numerous arc-shaped dislocation loop networks can be noticed (Fig. 9). Moreover, in the zirconia-ODS sample, multiple point defect clusters among dislocation can be observed. This difference between the type of crystal defects in unmodified material may be fundamental from the view of mechanisms occurring during irradiation. It is known that irradiation by charged particles such as protons, neutrons, or ions results in a progressive increase in the mean recoil energy. Consequently, non-uniform point defect generation occurs. Non-uniform character is strictly related to the production of displacement cascades by primary knock-out atoms [24,48]. During an ongoing cascade, the interstitials are transported outwards, displacing the last atom in the same crystallographic direction. Thus, the vacancy-rich region at the center of the cascade is created. High supersaturation of vacancies produces stress, which may be sufficiently large to create new dislocation. New dislocation provides an additional defect sink, where radiation-induced stresses may be reduced rapidly [48]. In the presented experiment, a similar effect may have occurred. Point defect clusters visible in zirconia-ODS specimens may have been transported outwards by progressing cascade effect during irradiation. At the same time, excess vacancies have generated stress sufficient for the creation of new dislocations. Generated dislocations reduce stress and provide extra sinks for the self-annealing process. Consequently, two phenomena may have occurred: (i) piling interstitials at the peak damage, where they may form interstitial loops typical for crystal metals [48], and (ii) generation of new dislocation below the peak damage as a result of stress generated by an excess of vacancies (as described above). This observation may be supported by relatively low microstrains calculated using GIXRD datasets (Tab. 3) – mentioned mechanisms lead to a sufficient reduction of generated stress and related microstrain of the lattice. In alumina-ODS samples, point defects are formed in a network of dislocation loops (Fig. 9). Hence, the redistribution mentioned above of the defects by the cascade and formation of new dislocations is less intensive than in zirconia-ODS specimens. Furthermore, it may be assumed that the mechanism of rapid reduction of stress does not occur (or is retarded). This is manifested by large microstrains estimated using GIXRD data (Tab. 3). Therefore, the estimated increment of dislocation density in the alumina-ODS specimen is much less intensive than in zirconia-ODS steel, while the inverse tendency is observed for microstrains' change. It must be highlighted that the TEM technique provides information from the limited volume of the material. Since some inhomogeneity of specimens has been observed (Fig. 8), the proposed explanation according to the presented TEM images should be taken carefully.

The irradiation temperature is another factor that strongly influences the mobility of radiation defects. The development of interstitials and vacancies generated from radiation damage is characterized by different kinetics at low and high temperatures. Moreover, as mentioned above, manufactured materials are characterized by high defect sinks. In this type of material, at low temperatures (as in the presented experiment), radiation-induced interstitials are more likely to diffuse to the nearest defect sink than interact with slower vacancies. Consequently, the concentration of vacancies continues to rise, whereas interstitials are loaded into the defect sinks. When a quasi-steady state of interstitial production and annihilation at sinks is reached, and the structure is supersaturated with vacancies, mutual recombination of radiation defects becomes dominant. Further generation of radiation defects leads to annihilating vacancies at defect sinks. The increased temperature leads to higher mobility of defects, which is crucial for vacancies. Additional thermal energy allows processes of defect annihilation (interstitials and vacancies) at defect sinks to occur faster than at low temperatures. Thus, mutual recombination at high temperatures has a minor effect on defects annihilation processes [24,49]. Moreover, it must be highlighted that the vacancy-diffusion mechanism is dominant in metal alloys [24]. Hence, diffusion in the structure is facilitated at high temperatures due to the higher concentration of vacancies and high grain boundaries volume.

Moreover, it is worth noticing that there are five temperature-dependent stages of defect recovery, as described in the work of Turos *et al.* [50]. Taking into account the temperature of the experiment (approx. 70°C) and the melting temperature of the studied material (1538°C for iron as a primary element [28]), stage III of recovery, where free migration of vacancies occurs, may be expected. The increased temperature of work (as in designs of IV generation of nuclear reactors) provides sufficient energy for the following stages that may take place ((IV) stage: development of vacancies clusters, and (V) dissociation of defect clusters and annealing of residual damage [50]). Furthermore, it has been proved that in metal alloys, the vacancy-diffusion mechanism is dominant, and grain boundaries play a role of high-diffusivity paths [24]. For the abovementioned reasons, one may expect that the radiation performance of alternative-oxides ODS steel would be more similar to yttria-ODS at a higher temperature.

## 5. Conclusions

In conclusion, four materials have been produced using mechanical alloying and Spark Plasma Sintering processes. Three materials have been strengthened by three different refractory oxides: zirconia, alumina, and - as a reference - commonly used yttria. Additional non-ODS reference material without any strengthening particles has been manufactured by using the same manufacturing route. Fabricated materials exhibit high relative density and are free from visible pores or voids. All studied materials have been exposed to high energy (500 keV) Ar-ion irradiation at room temperature. The process was done up to three fluences: $1 \times 10^{14}$, $1 \times 10^{15}$, and $5 \times 10^{15}$ ions/cm$^2$. This procedure resulted in the generation of a strongly modified thin layer with approximately a thickness of 230 nm with numerous radiation defects. Structural and nanomechanical investigation of radiation-affected layers has revealed a strong relationship between the type of introduced refractory oxide and the material's behavior under ion irradiation. It has been noticed that zirconia and alumina interact with the matrix elements and do not efficiently limit grain coarsening during the production process. As a consequence, ODS steels strengthened by alternative oxides do not exhibit as good performance under ion irradiation as reference yttria-ODS material. Nevertheless, the difference between materials up to the irradiation dose $1 \times 10^{15}$ ions/cm$^2$ is negligible. This suggests that mechanisms of self-annealing during radiation damage occur with similar efficiency in all ODS materials, independently from the type of introduced refractory oxide. Increasing the radiation dose leads to a strong differentiation between the materials. Steels strengthened by alternative oxides exhibit a much stronger hardening effect than yttria-ODS. Responses of zirconia- and alumina-ODS on radiation dose above $1 \times 10^{15}$ ions/cm$^2$ are similar to the response of non-ODS steel. Observed structural differences in visible defect forms between both materials may underlay dissimilarities in recorded results. Forms of pre-existing crystal defects determine the dominant mechanism occurring during radiation damage. These mechanisms significantly affect the increase of stress and microstrains estimated in the irradiated layer.

**Acknowledgments**

Financial support from National Science Centre, Poland through the PRELUDIUM 18 programme in the frame of grant no. 2019/35/N/ST5/00010 is gratefully acknowledged.